\documentclass[a4paper]{jpconf}

\usepackage{graphicx}
\usepackage{wasysym}
\usepackage{amssymb}
\usepackage{amsthm}

  \newcommand{\cB}{{\cal B}}

  \newcommand{\cP}{{\cal P}}
  
  \newcommand{\cT}{{\cal T}}
  \newcommand{\cV}{{\cal V}}

\begin{document}

\title{Anomalous transport and massive gravity theories}

\author{Eugenio Meg\'{\i}as$^{1,2}$}
 
\address{$^{1}$Max-Planck-Institut f\"ur Physik (Werner-Heisenberg-Institut), F\"ohringer Ring 6, D-80805, Munich, Germany}
\address{$^{2}$Departamento de F\'{\i}sica Te\'orica, Universidad del Pa\'{\i}s Vasco UPV/EHU, Apartado 644,  48080 Bilbao, Spain}

\ead{emegias@mppmu.mpg.de}

\begin{abstract}
We review the Kubo formulae relevant to study anomalous transport properties of relativistic fluids. We apply this formalism to perform a computation of the transport coefficients in a holographic massive gravity model including vorticity and external electromagnetic fields. We find an interesting phase in which the electric DC conductivity is negligible, while the anomalous conductivities turn out to be nonvanishing.
\end{abstract}

\section{Introduction}
\label{sec:intro}

Quantum anomalies are one of the most intriguing properties of relativistic field theories~\cite{Bertlmann:1996xk}. In the past few years it has been extensively studied the role played by quantum anomalies in the hydrodynamics of relativistic fluids. The hydrodynamical systems should obey the conservation laws of the energy-momentum tensor and spin one currents. In presence of anomalies the currents are no longer conserved, and this has important consequences in the constitutive relations for hydrodynamics. Two relevant phenomena induced by anomalies appear at first order in the hydrodynamical expansion: the {\it chiral magnetic effect}, which is the responsible for the generation of an electric current parallel to a magnetic field~\cite{Fukushima:2008xe}, and the {\it chiral vortical effect}, in which the electric current is induced by a vortex~\cite{Son:2009tf}. It is believed that these phenomena can produce observable effects in heavy ion physics~\cite{KerenZur:2010zw}, as well as in condensed matter systems~\cite{Basar:2013iaa}. It has been proposed in the literature several methods to compute the transport coefficients from a microscopic theory, either dissipative or non-dissipative, including kinetic theory~\cite{Arnold:2000dr}, Kubo formulae~\cite{Amado:2011zx,Landsteiner:2012kd,Chowdhury:2015pba}, diagrammatic methods~\cite{Manes:2012hf}, fluid/gravity correspondence~\cite{Bhattacharyya:2008jc,Erdmenger:2008rm,Banerjee:2008th}, and the partition function formalism~\cite{Banerjee:2012iz,Jensen:2012jy,Jensen:2012jh,Megias:2014mba}.

In addition to anomalies, there are other sources of non-conservation of the currents. In this context it is worth mentioning {\it disorder} as a very interesting property of some condensed matter systems~\cite{Lee:1985zzc}. Physical systems affected by disorder include real-world crystals with impurities, and the disorder phenomena can be manifested in the form of metal-insulator transitions~\cite{Imada:1998zz}. From a quantum field theory perspective, disorder has to do with explicit breaking of translation invariance, so that it leads to momentum dissipation and the conservation of the energy-momentum tensor is modified.  The study of disorder is particularly interesting in holography, as the AdS/CFT correspondence is a powerful tool that might help to establish some universal properties of these systems. Several holographic massive gravity models in AdS spacetimes have been recently proposed as a new and interesting approach to tackle this problem~\cite{Baggioli:2016oqk,Gouteraux:2016wxj}. It has been studied in these models some dissipative transport coefficients, like the DC electric conductivity, which turns out to be decreasing when disorder increases.

In this work we will review the Kubo formalism to study anomalous transport in hydrodynamics. In a second step, this formalism will be applied to study anomalous transport phenomena in a massive gravity model. We will contrast the result for the DC electric conductivity, with the one obtained for anomalous conductivities.

\section{Hydrodynamics of relativistic fluids}
\label{sec:hydro}

Hydrodynamics is a useful approach to study many phenomena for physical systems out of equilibrium. It can be applied when the mean free path of particles is much shorter than the characteristic size of the system. The basic ingredients are the constitutive relations, which are expressions of the energy-momentum tensor and the charge currents in terms of fluid quantities~\cite{kovtun:2012rj}, 
\begin{eqnarray}
\langle T^{\mu\nu} \rangle &=& (\varepsilon + P) u^\mu u^\nu  + P g^{\mu\nu} + \langle T^{\mu\nu} \rangle_{\textrm{\scriptsize diss \& anom}}   \,, \nonumber \\
\langle J^\mu \rangle &=& n u^\mu + \langle J^\mu \rangle_{\textrm{\scriptsize diss \& anom}}  \,. \nonumber
\end{eqnarray}
Here $\varepsilon$ is the energy density, $P$ the pressure, $n$ the charge density and $u^\mu$ the local fluid velocity. The equilibrium contributions are well known in the literature for many physical systems; however there are extra terms in the constitutive relations which lead to dissipative and anomalous effects. These extra contributions are in general difficult to compute, and the standard approach in hydrodynamics is to organize them in a derivative expansion, also called hydrodynamic expansion~\cite{Erdmenger:2008rm,Banerjee:2008th,Kharzeev:2011ds,Megias:2013joa}. In the Landau frame, defined as the one in which the energy flux vanishes at rest, the hydrodynamic expansion at first order in derivatives reads
\begin{eqnarray}
\langle T^{\mu\nu} \rangle_{\textrm{\scriptsize diss \& anom}} &=& - \eta P^{\mu\alpha}P^{\nu\beta} \left(D_\alpha u_\beta \!+\! D_\beta u_\alpha \!-\! \frac{2}{3}g_{\alpha\beta}D^\lambda u_\lambda\right) - \zeta P^{\mu\nu} D^\alpha u_\alpha + \cdots  \,, \label{eq:Tmunu} \\
\hspace{-0.9cm}\langle J^\mu \rangle_{\textrm{\scriptsize diss \& anom}}\!\!\!\! &=&\!\!\!\! -\sigma T P^{\mu\nu} D_\nu \left( \frac{\mu}{T} \right) + \sigma E^\mu + \sigma^\cB B^\mu + \sigma^\cV\omega^\mu + \cdots  \,, \label{eq:Jmu}
\end{eqnarray}
where $P^{\mu\nu} = g^{\mu\nu} + u^\mu u^\nu$ is the projector in the space orthogonal to the fluid velocity, $E^\mu = F^{\mu\nu} u_\nu$ and $B^\mu = \frac{1}{2}\epsilon^{\mu\nu\rho\lambda} u_\nu F_{\rho\lambda}$ are the electric and magnetic fields respectively, with the field strength of the gauge field defined as $F_{\mu\nu} = D_\mu A_\nu - D_\nu A_\mu$, and  $\omega^\mu = \epsilon^{\mu\nu\rho\lambda}u_\nu D_\rho u_\lambda$ is the vorticity. The shear~$\eta$ and bulk~$\zeta$ viscosities play a prominent role in heavy ion experiments, as they are dissipative coefficients that are responsible for most of the entropy production of the quark gluon plasma during the collision, see e.g.~\cite{Bozek:2009dw}. The electric conductivity $\sigma$ is also dissipative, and its effects is to generate an electric current in the medium induced by an external electric field. Finally, the coefficients $\sigma^\cB$ and $\sigma^\cV$ are the so-called chiral magnetic~\cite{Fukushima:2008xe} and chiral vortical~\cite{Son:2009tf} conductivities respectively, and they lead also to an electric current induced, either by an external magnetic field, or by a vortex in the fluid. As we will explain in Sec.~\ref{sec:Kubo_Formulae}, the coefficients $\sigma^\cB$ and $\sigma^\cV$ receive contributions from quantum anomalies, so that we will refer to them as anomalous conductivities.

In order to understand the (non)dissipative character of these transport coefficients, let us study their time reversal $\cT$ properties. The second law of thermodynamics states that the entropy should increase with time, i.e.
\begin{equation}
\frac{\partial }{\partial t} s > 0 \,. \label{eq:s}
\end{equation}
One can see from Eq.~(\ref{eq:s}) that only $\cT$-odd transport coefficients can contribute to the entropy production, while $\cT$-even coefficients cannot. On the one hand the electric current $\vec{J}$ is $\cT$-odd, while the electric field $\vec{E}$ is $\cT$-even, and both the magnetic field $\vec{B}$ and vorticity $\vec{\omega}$ are $\cT$-odd. Then, from a quick inspection of Eq.~(\ref{eq:Jmu}) one concludes that the electric conductivity $\sigma$ is $\cT$-odd, while the chiral magnetic and vortical conductivities, $\sigma^\cB$ and $\sigma^\cV$, are $\cT$-even. This explains the dissipative properties of $\sigma$, and the non-dissipative characters of $\sigma^\cB$ and $\sigma^\cV$. From a similar analysis of the parity $\cP$ properties one finds that the chiral conductivities are related to $\cP$-odd transport, see e.g.~\cite{Kharzeev:2011ds}.

It is possible to observe non-dissipative effects also in the energy-momentum tensor under an appropriate choice of the frame (see Sec.~\ref{subsec:strong_coupling}), in particular the generation of an energy-flux induced by a magnetic field or by a vortex, i.e.
\begin{eqnarray}
\langle T^{\mu\nu} \rangle_{\textrm{\scriptsize anom}} &=& \sigma_{\varepsilon}^{\cB} (B^\mu u^\nu + B^\nu u^\mu)  + \sigma_{\varepsilon}^{\cV} (\omega^\mu u^\nu +  \omega^\nu u^\mu) \,. \label{eq:Tanom}
\end{eqnarray}
We will refer to $\sigma_{\varepsilon}^{\cB}$ ($\sigma_{\varepsilon}^{\cV}$) as chiral magnetic (vortical) conductivity of energy current. In the rest of this work we will study the transport coefficients $\sigma$, $\sigma^\cB$, $\sigma^\cV$, $\sigma_{\varepsilon}^{\cB}$ and $\sigma_{\varepsilon}^{\cV}$. We will be especially focused on the chiral conductivities.

\section{Kubo Formulae}
\label{sec:Kubo_Formulae}

On the basis of linear response theory, hydrodynamic transport coefficients can be extracted from the long-wavelength and low-frequency limits of some retarded Green functions, leading to the so called Kubo formulae. The Kubo formula for the chiral magnetic conductivity was derived in~\cite{Kharzeev:2009pj}, while the one for the chiral vortical conductivity was studied in~\cite{Amado:2011zx}. They read, respectively, 
\begin{equation}
\sigma^\cB = \lim_{p_c\rightarrow 0} \frac{i}{2p_c} \sum_{a,b}\epsilon_{abc}
\langle J^a J^b \rangle|_{\omega=0}  \,,\quad \sigma^\cV  = \lim_{p_c\rightarrow 0} \frac{i}{2p_c} \sum_{a,b}\epsilon_{abc}
\langle J^a T^{0b} \rangle|_{\omega=0} \,.  \label{eq:Kubo}
\end{equation}
Similar Kubo formulae have been derived for the transport coefficients related to the generation of energy flux, in particular $\sigma_{\varepsilon}^{\cB} \sim \epsilon_{abc} \langle T^{0a} J^b \rangle$ and $\sigma_{\varepsilon}^{\cV} \sim \epsilon_{abc} \langle T^{0a} T^{0b} \rangle$, see~\cite{Amado:2011zx,Landsteiner:2011iq,Landsteiner:2012kd,Chowdhury:2015pba}. We will study in this section the result of the computation of Eq.~(\ref{eq:Kubo}) at weak and strong coupling.

\subsection{Weak coupling results}
\label{subsec:weak_coupling}

Let us consider a theory of $N$ free chiral fermions transforming under a global symmetry group $G$ generated by matrices $(T_A)^f\,{}_g$. The chemical potential for the fermion $\Psi^f$ is given by $\mu^f= \sum_A q_A^f \mu_A$, while the Cartan generator is $H_A = q^f_A \delta^f\,{}_g$ where $q^f_A$ are the charges. It has been performed in the literature the 1-loop computation of the chiral magnetic~\cite{Kharzeev:2009pj} and the chiral vortical~\cite{Landsteiner:2011cp} conductivities by using the Kubo formulae of Eq.~(\ref{eq:Kubo}). The results read, respectively,
\begin{equation}
  (\sigma^\cB)_{AB} = \frac{1}{4\pi^2} d_{ABC} \mu^C  \,, \qquad (\sigma^\cV)_A = \frac{1}{8\pi^2}  \sum_{B,C} d_{ABC} \, \mu^B \, \mu^C  + \frac{T^2}{24} b_A \,, \label{eq:sigmaBV}
\end{equation}
where the coefficients
\begin{equation}
 d_{ABC} = \frac{1}{2} \left[ \tr( T_A \{ T_B, T_C \})_R - \tr( T_A \{ T_B, T_C \})_L \right] \,, \qquad   b_A =  \tr \left( T_A \right)_R -  \tr \left( T_A \right)_L   \,, \label{eq:db}
\end{equation}
are related to the trace of the generators of the symmetry group. The subscripts $R$, $L$ stand for the contributions of right-handed and left-handed fermions.  One can easily identify $d_{ABC}$ with the group theoretic factor related to the axial anomaly, which typically appears in the computation of the anomalous triangle diagram corresponding to three non-abelian gauge fields coupled to a chiral fermion.  On the other hand, the coefficient $b_A$ corresponds to the mixed gauge-gravitational anomaly~\cite{Kumura:1969wj}, appearing in the anomalous triangle diagram with one non-abelian gauge field and two insertions of the energy-momentum tensor coupled to a chiral fermion. This means that the chiral magnetic and chiral vortical conductivities are induced by chiral anomalies, either the axial anomaly or the mixed gauge-gravitational anomaly.

Anomalies are also responsible for a non-vanishing value of the divergence of the currents, which in this case reads~\cite{AlvarezGaume:1983ig}~\footnote{The gauge-gravitational anomaly is the statement that it is not possible to preserve at the same time the vanishing of the divergence of the energy-momentum tensor and of the chiral $U(1)$ currents.}
\begin{eqnarray}
&&D_\mu J_A^\mu= \epsilon^{\mu\nu\rho\lambda}\left( \frac{d_{ABC}}{32\pi^2} F^B_{\mu\nu} F^C_{\rho\lambda} + \frac{b_A}{768\pi^2} 
R^\alpha\,_{\beta\mu\nu} R^\beta\,_{\alpha\rho\lambda}  \right) \,.  \label{eq:DJmu}
\end{eqnarray}
A clear consequence of Eq.~(\ref{eq:DJmu}) is that anomalies modify the hydrodynamic equations, due to the non-conservation of the currents. So it is obvious that they should affect the hydrodynamic expansion with new contributions not appearing in absence of anomalies. We have identified above these new contributions at first order in derivatives as $\sigma^\cB$ and $\sigma^\cV$ in~Eq.~(\ref{eq:Jmu}).~\footnote{$\sigma_{\varepsilon}^{\cB}$ and $\sigma_{\varepsilon}^{\cV}$ in Eq.~(\ref{eq:Tanom}) are induced also by chiral anomalies. A computation of these conductivities at weak coupling in line with this section can be found in e.g.~\cite{Landsteiner:2012kd,Landsteiner:2013aba}.}

\subsection{Strong coupling results}
\label{subsec:strong_coupling}

The Kubo formulae Eq.~(\ref{eq:Kubo}) have been computed in~\cite{Landsteiner:2011iq,Megias:2013joa,Landsteiner:2011tf} at strong coupling within a Einstein-Maxwell model in 5 dim. In order to mimic the anomalous effects, the model is supplemented with pure gauge and mixed gauge-gravitational Chern-Simons terms. The action reads
\begin{eqnarray}
 S &=& \frac{1}{16\pi G} \int d^5x \sqrt{-g} \Bigg[ R + 12 - \frac 1 4 F_{MN} F^{MN} \nonumber \\ 
&&+ \epsilon^{MNPQR} A_M \left( \frac\kappa 3 F_{NP} F_{QR} + \lambda R^A\,_{BNP} R^B\,_{AQR}   \right) \Bigg] + S_{GH} + S_{CSK} \,, \label{eq:model1}
\end{eqnarray}
where $S_{GH}$ is the usual Gibbons-Hawking boundary term, and $S_{CSK}$ is an extra boundary contribution needed to reproduce the mixed gauge-gravitational anomaly at a general hypersurface. A computation of the current with this model leads to
\begin{equation}
\tilde{J}^\mu = \frac{\partial S}{\partial A_\mu} = -\frac{\sqrt{-\gamma}}{16\pi G} \left[ F^{r\mu} + \frac{4}{3}\kappa \epsilon^{\mu\nu\rho\lambda} A_\nu F_{\rho\lambda}\right]_\partial \equiv J^\mu + K^\mu \,. \label{eq:Jconsistent}
\end{equation}
Note that $\tilde{J}^\mu$ is not gauge covariant, and we refer to it as {\it consistent} current. The {\it covariant} version of the current, denoted by $J^\mu$, is the usual one appearing in the constitutive relations. From Eq.~(\ref{eq:Jconsistent}), an on-shell computation of the divergence of the covariant current leads to the anomaly~\footnote{Quantities with hat $({\hat F}, {\hat R}, \cdots)$ refer to their induced four dimensional objects at a cut-off surface.}
\begin{equation}
D_\mu J^\mu = -\frac{1}{16\pi G} \epsilon^{\mu\nu\rho\lambda} \left(
\kappa {\hat F}_{\mu\nu} {\hat F}_{\rho\lambda} + \lambda {\hat R}^\alpha\,_{\beta\mu\nu} {\hat R}^\beta\,_{\alpha\rho\lambda} \right) \,.
\end{equation}
Finally, from a comparison with Eq.~(\ref{eq:DJmu}) one can identify the parameters $\kappa$ and $\lambda$ as the axial anomaly and mixed gauge-gravitational anomaly coefficients, respectively.

The holographic computation of the transport coefficients with the Kubo formulae follows from the study of retarded propagators by using the AdS/CFT dictionary~\cite{Son:2002sd,Herzog:2002pc}. In general lines, we split the metric and gauge field into a background and a linear perturbation
\begin{equation}
g_{MN} = g^{(0)}_{MN} + \epsilon \, h_{MN} \,, \qquad A_M = A_M^{(0)} + \epsilon \, a_M \,.
\end{equation}
Inserting these fields in the action and expanding up to second order in $\epsilon$ one can obtain the second order action, and from there the desired propagators~\cite{Kaminski:2009dh}. The equations of motion of the fluctuations should obey appropriate boundary conditions, in particular: i) regularity at the horizon; and ii) vanishing at the boundary, as the fluctuations cannot modify the boundary values, which are given by the background. There is some freedom in one of the boundary conditions which is related to the choice of frame. This corresponds to the particular definition of the local fluid velocity.

We show in Table~\ref{tab:coef} the values of the anomalous conductivities computed with the holographic model of Eq.~(\ref{eq:model1}), corresponding to a system which realizes a single $U_A(1)$ symmetry. We display the results in three different frames. $u^\mu$ can be taken to be proportional to: i) the energy flux $\langle T^{0i} \rangle = (\varepsilon + P) u^i$ (Landau frame), and ii) the charge current $\langle J^i \rangle = n u^i$ (Eckart frame). In addition, the laboratory rest frame is the natural frame in the field theory computation, see Sec.~\ref{subsec:weak_coupling}, and it seems to be related to the entropy current~$\langle J^i_S \rangle_{\textrm{\scriptsize anom}} = 0$, see e.g.~\cite{Megias:2013xla}.

\begin{table}[h]
\begin{center}
\begin{tabular}{|l|l|l|l|}
\hline
Conductivities & Laboratory rest frame &  Landau frame & Eckart frame  \\
\hline\hline
 $\qquad (\sigma^\cB)_\textrm{\tiny F}$ & $\sigma^\cB =  \frac{ \mu}{4 \pi^2}$   & $\sigma^\cB - \frac{n}{\varepsilon + P} \sigma_{\varepsilon}^{\cB}$    &   $\qquad 0$   \\ 
 $\qquad (\sigma^\cV)_\textrm{\tiny F}$ & $\sigma^\cV = \frac{\mu^2}{8\pi^2} + \frac{T^2}{24} $   &  $\sigma^\cV - \frac{n}{\varepsilon + P} \sigma_{\varepsilon}^{\cV}$   & $\qquad 0$   \\
 $\qquad (\sigma_{\varepsilon}^{\cB})_\textrm{\tiny F}$   &  $\sigma_{\varepsilon}^{\cB} = \sigma^\cV$   &  $\qquad 0$     & $\sigma_{\varepsilon}^{\cB} -\frac{\varepsilon + P}{n}\sigma^\cB$     \\
$\qquad (\sigma_{\varepsilon}^{\cV})_\textrm{\tiny F}$  &   $\sigma_{\varepsilon}^{\cV} = \frac{\mu^3}{12 \pi^2} + \frac{\mu T^2}{12}$     &  $\qquad 0$     &  $ \sigma_{\varepsilon}^{\cV} -\frac{\varepsilon+P}{n}\sigma^\cV $      \\
\hline
\end{tabular}
\caption{\label{tab:coef} Anomalous conductivities contributing to the constitutive relations, Eqs.~(\ref{eq:Jmu}) and (\ref{eq:Tanom}), for a theory with a single chiral $U_A(1)$ symmetry. The result at weak and strong coupling agree.}
\end{center}
\end{table}

Let us finally stress that the values of the anomalous conductivities at strong coupling coincide precisely with the ones obtained at weak coupling, and this a strong hint towards a non-renormalization theorem for the anomalous conductivities, see e.g.~\cite{Landsteiner:2011cp,Landsteiner:2011iq,Landsteiner:2011tf}.~\footnote{Note, however, that in presence of dynamical gluons some renormalization effects have been observed in $\sigma^\cV$, see~\cite{Hou:2012xg,Jensen:2012kj,Jensen:2013vta,Golkar:2012kb}.}

\section{Transport properties in massive gravity}
\label{sec:massive_gravity}

It has been recently studied in~\cite{Baggioli:2016oqk,Gouteraux:2016wxj} some minimal models for massive gravity in 4 dim. The momentum relaxation is described through the Stueckelberg mechanism with Goldstone modes corresponding to scalars $X^I$, which are related to spatial translations. In this section we will consider the model presented in~\cite{Gouteraux:2016wxj}. We have explicitly checked that our conclusions are the same when considering the model of~\cite{Baggioli:2016oqk}.

\subsection{The model}
\label{subsec:model}

In order to study anomalous transport in holographic massive gravity theories, one should consider the theory in odd dimensions, as only in these cases there is contribution from the chiral anomaly and we can introduce the corresponding Chern-Simons terms, see Eq.~(\ref{eq:model1}). We will consider the model of~\cite{Gouteraux:2016wxj}, and extend it to 5 dim. The action reads
\begin{equation}
 S = \frac{1}{16\pi G} \int d^5x \sqrt{-g} \left[ R + 12 - \frac 1 2 \partial^M X^I \partial_M X^I - \frac 1 4 F^2 - \frac {\cal J} 4 \partial_M X^I \partial_N X^I F^N\,_L F^{L M}  \right] \,, \label{eq:massive_gravity}
\end{equation}
with scalar fields $X^I = k \delta^I_i x^i$, where $i$ denotes spatial directions. Momentum relaxation is implemented by giving the scalar fields a vacuum expectation value. The term $\sim {\cal J}$ is a higher derivative coupling between the charge and the scalar field. The parameter $k$ controls the degree of breaking of translational invariance, in particular when $k=0$ one recovers the massless gravity theory. The equations of motion for the fields $(\delta g^{MN}, \delta A^M, \delta X^I)$ read
\begin{eqnarray}
 0 &=& G_{MN}-6 g_{MN} + \frac{1}{2} F_{M}{}\,^L F_{L N} - \frac{1}{8} g_{MN} F^2 - \frac{1}{2} \partial_M X^I\partial_N X^I +
\frac 1 4 g_{MN} (\partial_L X^I)(\partial^L X^I) \nonumber \\
&&- \frac {\cal J} 4 \left(\tilde X.F.F+F.\tilde X.F+ F.F\tilde X \right)_{MN} 
+ \frac {\cal J} 8 g_{MN} \mathrm{tr}(\tilde X.F.F) \,, \label{eq:eomg} \\
0 &=& \nabla_N F^{NM} + \frac {\cal J} 2 \nabla_L(\tilde X . F)^{ML} - \frac {\cal J} 2 \nabla_L(\tilde X. F)^{LM} \,,  \label{eq:eomA}\\
0 &=& \Box X^I + \frac{{\cal J}}{2} \nabla_M ( \partial_N X^I F^N\,_L F^{LM})\,,  \label{eq:eomX}
\end{eqnarray}
where $\tilde X_{MN}  \equiv \partial_M X^I \partial_N X^I$, and the products are $\tilde X.F.F =  \partial_M X^I \partial_N X^I F^N\,_L F^{LM}$, etc. These equations can be solved with the black-brane ansatz:
\begin{equation}
 ds^2  = -f dt^2 + \frac{dr^2}{f} + r^2 d\vec x^2 \,, \qquad A = A_t \, dt \,.
\end{equation}
We find the following solution:
\begin{equation}
f = r^2 - \frac{M}{r^2}  + \frac{\rho^2}{3 r^4} + \frac{k^2}{2} \,, \qquad A_t = \mu - \frac{\rho}{r^2} \,,
\end{equation}
and the temperature and chemical potential can be related to the mass $M$ and charge $\rho$ of the black hole as
\begin{equation}
 T = \frac{1}{2\pi r_h}\left( \frac{2M}{r_h^2} -\mu^2 - \frac{k^2}{2} \right) \,, \qquad \mu r_h^2 = \rho \,,
\end{equation}
where $r_h$ is the black hole horizon corresponding to the largest solution of $f=0$.

\subsection{Electric DC conductivity}
\label{subsec:DC_conductivity}

\begin{figure*}[htb]
\begin{tabular}{cc}
\includegraphics[width=75mm]{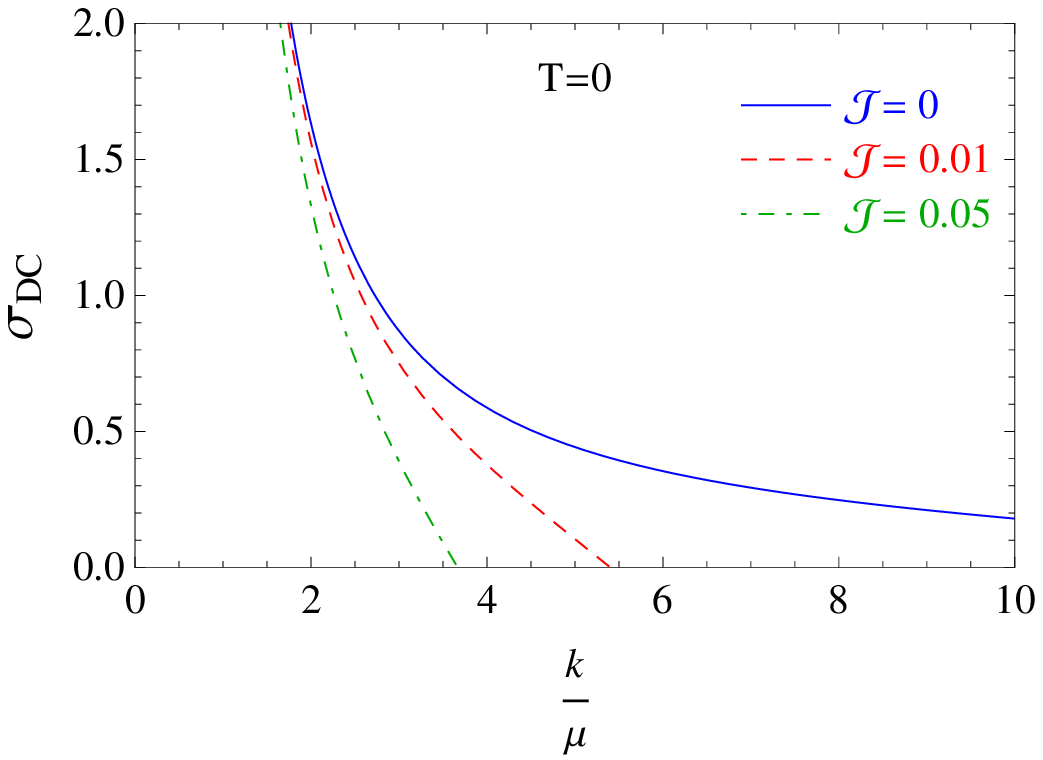} &
\includegraphics[width=75mm]{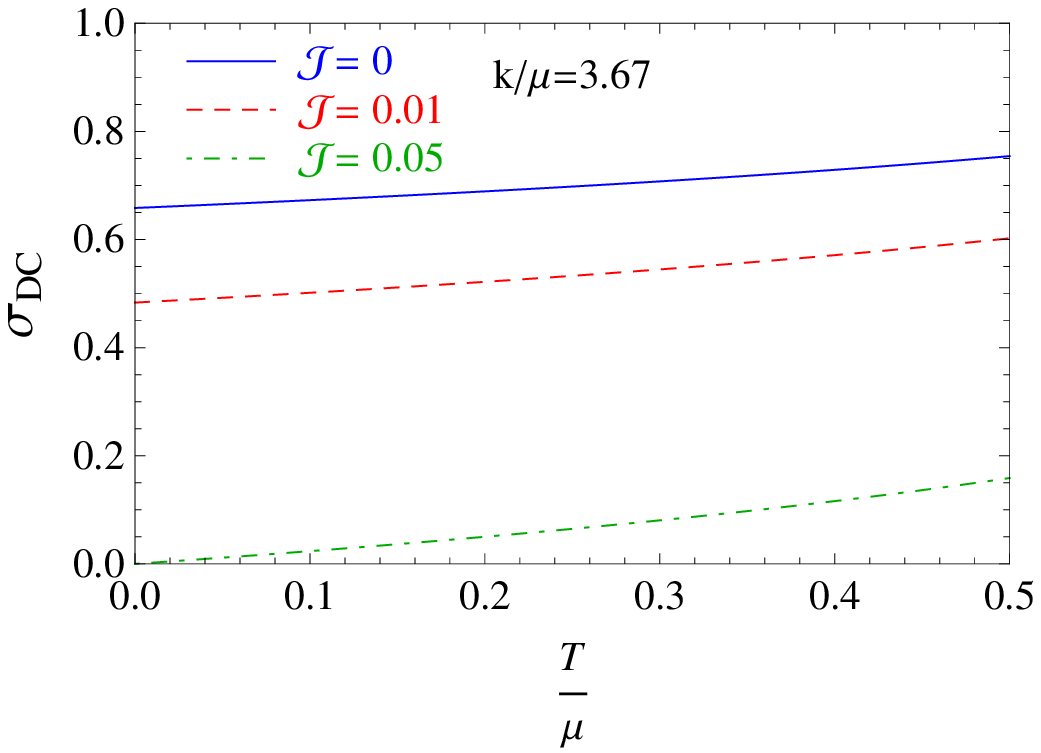} \\
 (A)
    &
         (B)
\end{tabular}
\caption{(A) DC conductivity at zero temperature as a function of $k$ (normalized to the chemical potential $\mu$). (B)~DC conductivity (at fixed $k/\mu=3.67$) as a function of temperature (normalized to $\mu$). These results are obtained with Eq.~(\ref{eq:DCconductivity}) for different values of the parameter~${\cal J}$.}
\label{fig:DCconductivity}
\end{figure*}

\begin{figure}[htb]
\includegraphics[width=75mm]{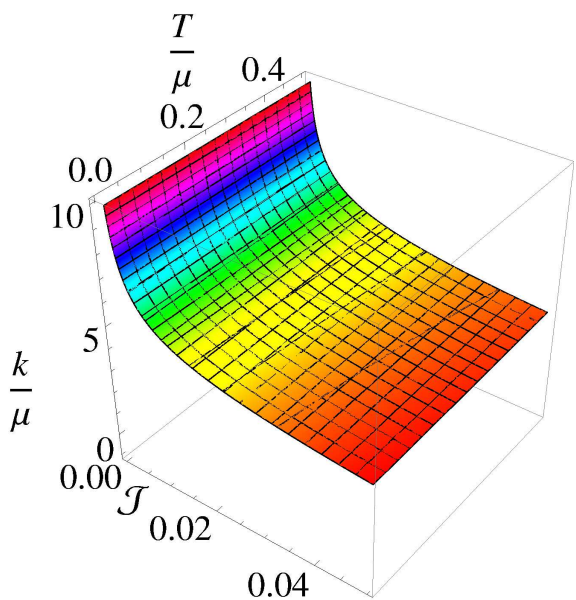}
\begin{minipage}{18pc}
\vspace{-3.5cm}\caption{Region in the plane $({\cal J},T/\mu,k/\mu)$ where the DC conductivity of Eq.~(\ref{eq:DCconductivity}) vanishes.}
\label{fig:DCconductivity3D}
\end{minipage}
\end{figure}

As shown in Eq.~(\ref{eq:Jmu}), the electric conductivity measures the electric current $J^\mu$ induced by an electric field $E^\mu$. There are several methods to compute the DC conductivity in holography, but one of the most straightforward is the one proposed in~\cite{Donos:2014cya}. We consider the following ansatz for the fluctuations in the gauge field and the metric:
\begin{equation}
A_x = \epsilon(E t + a_x(r))  \,, \qquad g_{tx} = r^2 (1+ \epsilon h_{tx}(r)) \,, \qquad g_{rx} = r^2 (1+ \epsilon h_{rx}(r)) \,,
\end{equation}
and work to first order in $\epsilon$. The fluctuation equations can be written now as
\begin{eqnarray}
 0= & \frac{k^2}{2} \left( 1 - \frac{2 {\cal J} \rho^2}{r^6} \right) h_{rx} - \left( 1 - \frac{{\cal J} k^2}{2r^2} \right)\frac{E \rho}{f r^3} \,, \label{eq:condhrx} \\
 r^8 f( h_{tx}'' + \frac{5}{r} h_{tx}' ) &= k^2( r^6 + 2 {\cal J} \rho^2 ) - \rho ( 2 r^2 - k^2 {\cal J}^2) (r f a_x')\,, \label{eq:condhtx} \\
 0= & \left[ \left(1 - \frac{{\cal J} k^2}{2 r^2} \right) r f a_x' + \left(2 - \frac{{\cal J} k^2}{r^2}\right) \rho h_{tx} \right]' \,.  \label{eq:condax}
\end{eqnarray}
The DC conductivity can be obtained from a solution of these equations with appropriate boundary conditions at the horizon and the boundary, as explained in Sec.~\ref{subsec:strong_coupling}. At this point we will skip the technical details, that will be presented in~\cite{workinprogress}. The final result for the DC conductivity reads
\begin{equation}
\sigma_{DC} = \frac{J_x}{E}= r_h \left(1 - \frac{{\cal J} k^2}{2 r_h^2}  \right) \left[1 + \left(1 - \frac{{\cal J} k^2}{2 r_h^2}  \right) \frac{ 4 \mu^2}{k^2 M(r_h)} \right] \,, \label{eq:DCconductivity}
\end{equation}
with $M(r_h) = 1 + \frac{2 {\cal J} \rho^2}{ r_h^6}$. This formula is the equivalent to the result presented in Eq.~(3.15) of~\cite{Gouteraux:2016wxj}, but in 5 dim. 
We plot in Fig.~\ref{fig:DCconductivity} the value of the DC conductivity from Eq.~(\ref{eq:DCconductivity}) as a function of the parameter $k$ (left) and the temperature $T$ (right) for different values of the parameter ${\cal J}$. On the one hand, we observe that in the case ${\cal J} = 0$, the conductivity is bounded from below, so that there is no insulating behavior without ${\cal J}$. More important is the fact that when ${\cal J} \ne 0$, there is a particular value $k_* = k({\cal J})$ at which the DC conductivity vanishes, so that the theory behaves as an insulator. This property is one of the main results of~\cite{Baggioli:2016oqk,Gouteraux:2016wxj} in 4 dim, and here we have reproduced it as well in 5 dim. Finally, we show in Fig.~\ref{fig:DCconductivity3D} the regime of parameters where the DC conductivity vanishes.

\subsection{Anomalous Transport}
\label{subsec:anomalous_massive_gravity}

In Sec.~\ref{subsec:DC_conductivity} we have studied some dissipative properties of the 5 dim massive gravity model. In order to study non-dissipative transport coefficients, we should introduce anomalous effects in the theory. As we explained in Sec.~\ref{sec:Kubo_Formulae}, anomalies are mimicked in the gravity side through Chern-Simons terms. When extending the model of Eq.~(\ref{eq:massive_gravity}) with the Chern-Simons terms of Eq.~(\ref{eq:model1}), extra contributions appear in the equations of motion of the theory, in particular the term $- 2 \lambda \epsilon_{LPQR(M} \nabla_B \left( F^{PL}R^B\,_{N)}\,^{QR}\right)$ in the rhs of Eq.~(\ref{eq:eomg}), and the terms $\epsilon^{MNPQR} (\kappa F_{NP} F_{QR} + \lambda R^A\,_{BNP} R^B\,_{AQR})$ in the rhs of Eq.~(\ref{eq:eomA}). By using the following conventions
\begin{equation}
ds^2 = \frac{du^2}{4 u^2 f(u)} + \frac{1}{u} \left( f(u) dt^2 + 2 h_{ti}(u) e^{i p z} dt dx^i + d\vec{x}^2 \right) \,,
\end{equation}
where $i=x,y$, the fluctuation equations are
\begin{eqnarray}
0 &=& (a_i' f)' - \rho h_{ti}'  - i p \epsilon_{ij} \left[ 4 \kappa \rho a_j + \lambda\left(12 u^2 \rho^2 + \frac{24}{u} (1-f) + 5k^2 \right) h_{tj}' \right]  \,, \label{eq:fluca}\\
0 &=& h_{ti}'' -\frac 1 u h_{ti}' -\frac{k^2}{4 u f} h_{ti} - u \rho a_i' + i p \lambda \epsilon_{ij} \left[\left(24(1-f) -12 u^3 \rho^2 - 5 u k^2 \right) a_j' + \right.\nonumber \\
 && + \left. \left( \frac{24}{u} (1-f) - 32 u^2 \rho^2 - 6k^2\right) a_j - 8 \rho u (u h_{ti}')' \right] \,.
\end{eqnarray}
Finally, from the procedure of Sec.~\ref{sec:Kubo_Formulae} we get the result for the anomalous conductivities:~\footnote{We show in these expressions the explicit dependence on the parameters $\kappa$ and $\lambda$ for an easier identification of the axial and gauge-gravitational anomaly contributions. One can compare with the results of Table~\ref{tab:coef} by considering $\kappa=1/(16\pi^2)$ and $\lambda= 1/(384\pi^2)$ in Eqs.~(\ref{eq:sB})-(\ref{eq:sVe}).}
\begin{eqnarray}
 \sigma^\cB &=&  \lim_{p\to 0} \frac{i}{p} \langle J^x J^y\rangle = \kappa 4\mu  \,, \label{eq:sB} \\
\sigma^\cV &=&   \lim_{p\to 0}  \frac{i}{p} \langle J^x T^{0y} \rangle = \kappa 2\mu^2 + \lambda 16\pi^2 T^2  \,, \label{eq:sV}  \\
 \sigma_\varepsilon^\cB &=&  \lim_{p\to 0}  \frac{i}{p} \langle T^{0x} J^y \rangle =  \kappa 2\mu^2 +\lambda 16 \pi^2 T^2 + \lambda 2 k^2 \,, \label{eq:sBe} \\
\sigma_\varepsilon^\cV &=&  \lim_{p\to 0}  \frac{i}{p} \langle T^{0x} T^{0y} \rangle = \kappa \frac{4}{3}\mu^3 + \lambda 32\pi^2 T^2 \mu \,. \label{eq:sVe}
\end{eqnarray}
Some remarks deserve to be mentioned at this point. On the one hand the chiral magnetic and vortical conductivities of charge currents, $\sigma^\cB$ and $\sigma^\cV$, and the chiral vortical conductivity of energy current, $\sigma_\varepsilon^\cV$, are the same as in the massless gravity theory, see Table~\ref{tab:coef}. This means that these anomalous conductivities are not affected by translational breaking effects, and this constitutes one of the most important results of this work. From this property, together with the result of Sec.~\ref{subsec:DC_conductivity}, we conclude that there is a regime in the theory in which the DC conductivity vanishes, but the anomalous conductivities do not vanish. A consequence is that one can study the anomalous effects of these systems in this regime in a clean way. Let us finally mention that, contrary to the other coefficients, the chiral magnetic effect of energy current, $\sigma_\varepsilon^\cB$, has some dependence on $k$. The interpretation of this result is currently in progress.~\footnote{Several hypotheses might be proposed for the dependence of $\sigma_\varepsilon^\cB$ on the parameter $k$, including renormalization or gauge artefacts, as well as lack of unitarity of the theory. In addition, one cannot exclude the possibility that this might be due to a pathology of the model. In fact, it has been studied in the literature the thermodynamics of some massive gravity theories, and an intriguing result has been found related to the non-vanishing value of the entropy in the ground state ($T=0$, $\mu=0$), see~\cite{Vegh:2013sk}.}

\section{Discussion and conclusions}
\label{sec:conclusions}

In this work we have reviewed the role played by the chiral anomalies in the hydrodynamics of relativistic fluids. In particular, we have focused on the effects of external magnetic fields and vortices in the fluid, and characterized how the chiral anomalies contribute to the corresponding conductivities through the anomaly coefficients, Eq.~(\ref{eq:db}). The anomalous conductivities turn out to be non-dissipative at first order in the hydrodynamic expansion.

Massive gravity models have been introduced in the literature as holographic duals of disorder in condensed matter systems. By using the Kubo formalism, we have computed the transport properties in a massive gravity model with higher derivative corrections. We found an interesting regime in which the electric DC conductivity vanishes, but the anomalous conductivities turn out to be nonvanishing. Moreover, the anomalous transport seem to be unaffected by translational breaking effects. These and other issues will be addressed in more detail in a forthcoming publication~\cite{workinprogress}.

\ack
I would like to thank M. Baggioli, S.D.~Chowdhury, J.R.~David, K. Jensen, O.~Pujol\'as and especially K.~Landsteiner for valuable discussions. I thank the Instituto de F\'{\i}sica Te\'orica UAM/CSIC, Madrid, Spain, for their hospitality during the completion of the final stages of this work. Research supported by the European Union under a Marie Curie Intra-European fellowship (FP7-PEOPLE-2013-IEF) with project number PIEF-GA-2013-623006. 

\section*{References}

\bibliography{Megias}

\end{document}